\documentclass[9pt,twocolumn,twoside]{osajnl}
\pdfoutput=1
\journal{ol}
\usepackage{upgreek}
\usepackage[english]{babel}
\usepackage[utf8]{inputenc}
\usepackage{float}
\usepackage{subcaption}
\usepackage{hhline}
\usepackage{array}
\newcolumntype{?}{!{\vrule width 1pt}}
\usepackage{amsmath}
\setboolean{shortarticle}{true}

\title{Better magneto-optical filters  with cascaded vapor cells in the Faraday-Faraday and Faraday-Voigt geometries.}

\author[1,*]{Fraser D. Logue}
\author[1]{Jack D. Briscoe}
\author[1]{Danielle Pizzey}
\author[1]{Steven A. Wrathmall}
\author[1]{Ifan G. Hughes}

\affil[1]{Department of Physics, Durham University, Durham, DH1 3LE, United Kingdom}

\affil[*]{Corresponding author: fraser.d.logue2@durham.ac.uk}

\begin{abstract}
Single-cell magneto-optical Faraday filters find great utility and are realized with either ‘wing’ or ‘line center’ spectral profiles. We show that cascading a second cell with independent axial (Faraday) or transverse (Voigt) magnetic field leads to improved performance in terms of figure of merit (FOM) and spectral profile. The first cell optically rotates the plane of polarization of light creating the high transmission window; the second cell selectively absorbs the light eliminating unwanted transmission. Using naturally-abundant Rb vapor cells, we realize a Faraday-Faraday wing filter and the first recorded Faraday-Voigt line center filter which show excellent agreement with theory. The two filters have FOM values of 0.86 and 1.63~GHz$^{-1}$ respectively, the latter of which is the largest FOM atomic line filter recorded.\\  
\end{abstract}
\setboolean{displaycopyright}{true}

\begin{document}
\maketitle
\begin{figure*}[tb]
\centering
{\includegraphics[width=0.5\linewidth]{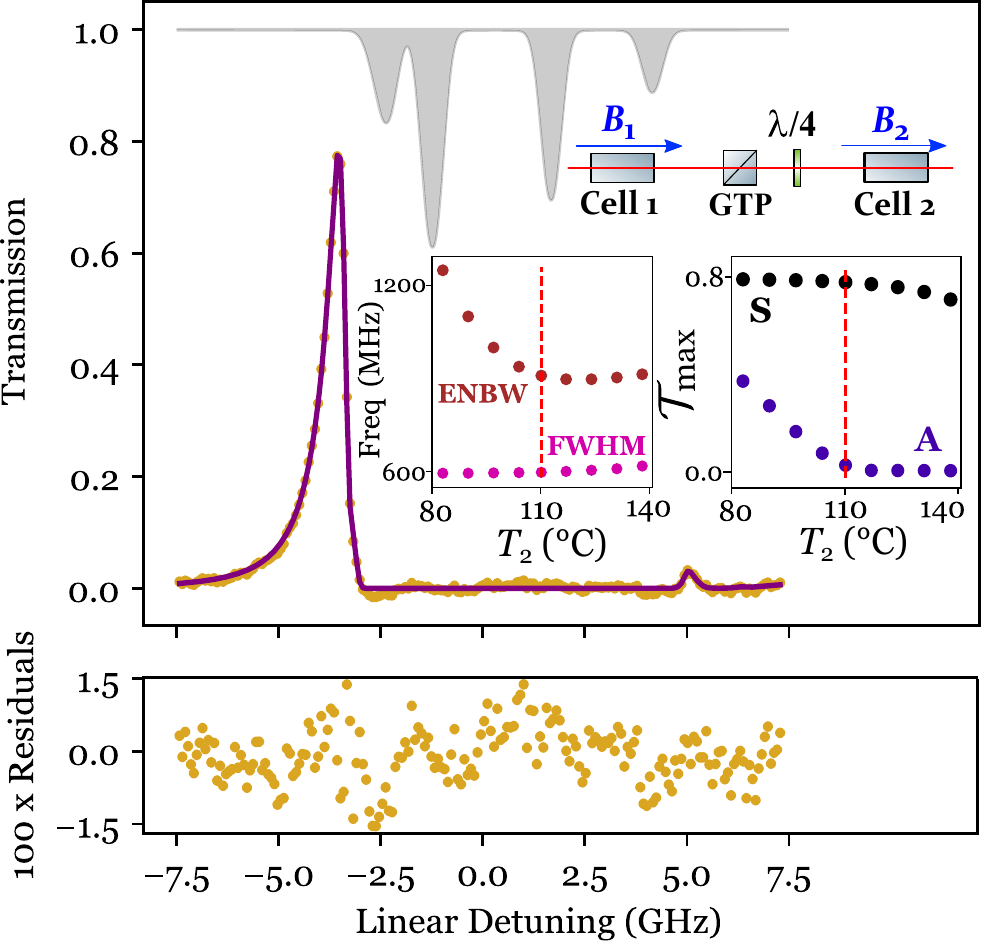}}\hspace{0.01cm} \hfill
{\includegraphics[width=0.49\linewidth]{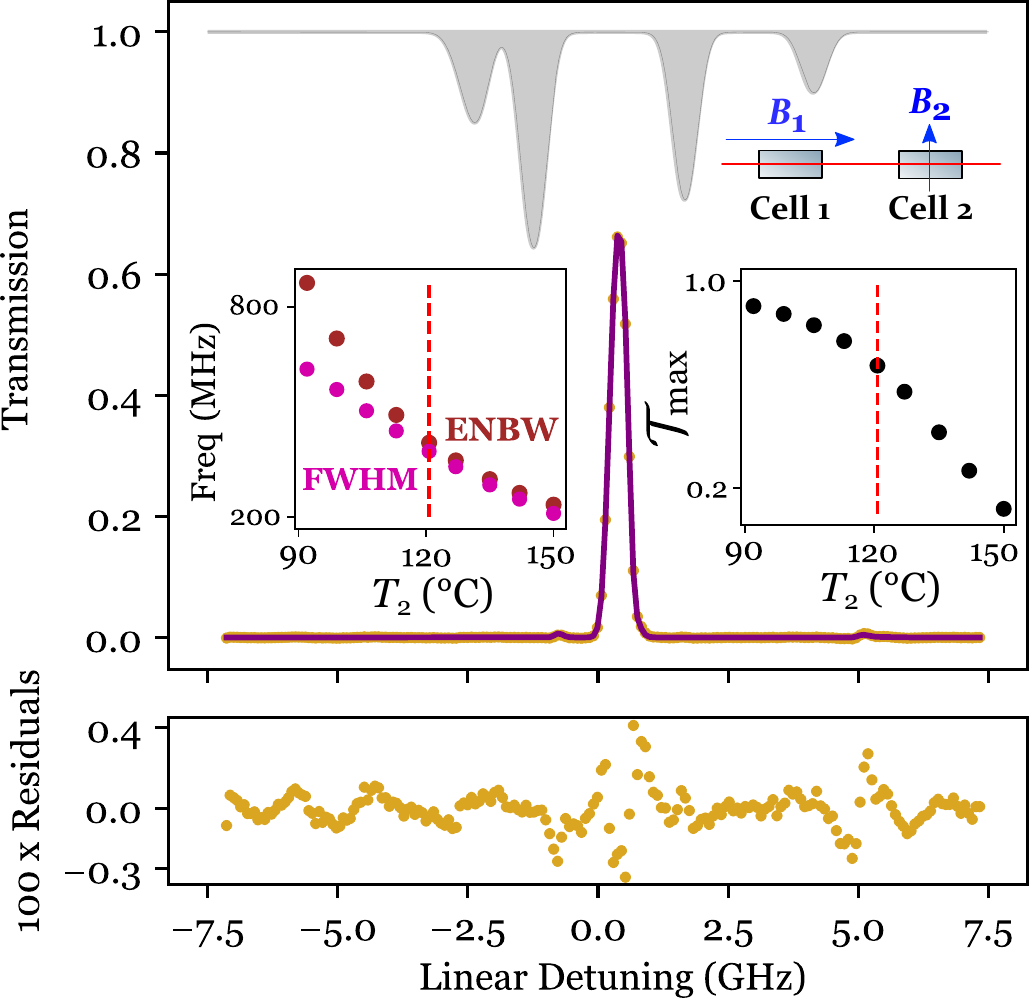}}
\begin{center}
\begin{tabular*}{1.5\linewidth}{|c|c|c|c|c|c|c|c|}
\cline{1-8}
\textbf{Filter Type} & $\boldsymbol{T_1\hspace{0.1cm}(^\circ\rm{C})}$ & $\boldsymbol{B_1\hspace{0.1cm}(\rm{G}) }$ & $\boldsymbol{T_2\hspace{0.1cm}(^\circ\rm{C})}$ & $\boldsymbol{B_2\hspace{0.1cm}(\rm{G}})$ & \textbf{ENBW} $\boldsymbol{(\rm{GHz})}$ & \textbf{FWHM} $\boldsymbol{(\rm{MHz})}$ & \textbf{FOM} $\boldsymbol{(\rm{GHz^{-1}})}$ \\ \hhline{|=|=|=|=|=|=|=|=|}
Wing & $85.6~\pm~0.3$ & $48.9 \pm 0.5$ & $110.1~\pm~0.3$ & $747 \pm 7$ & $0.92 ~\pm ~0.01$ & $599 ~\pm ~1$ & $0.86 ~\pm ~0.01$ \\ \cline{1-8}
Line Center & $100.29~ \pm~ 0.07$ & $162.2~ \pm~ 0.3$ & $120.79~ \pm~ 0.08$ & $2527.6~ \pm~ 0.3$ & $0.42~ \pm~ 0.01$ & $389~ \pm~ 1$ & $1.63~ \pm~ 0.01$ \\ \cline{1-8}
\end{tabular*}
\end{center}
\caption{(Left) Main plot shows data (gold) and theory (purple) plotted for a natural abundance Rb-D2 Faraday-Faraday wing filter with 75~mm and 5~mm cells. (Right) Main plot shows data (gold) and theory (purple) plotted for a natural abundance Rb-D2 Faraday-Voigt line center filter with two 5~mm cells. Both sets of data show excellent agreement with theory with RMS fit errors of 0.6\% and 0.09\% respectively. The insets show theory plots of ENBW, FWHM and maximum transmission against second cell temperature. On the left plot, the maximum transmission of both the selected (S) peak and the suppressed (A) peak are plotted. All other parameters are fixed. The red dotted line indicates the experimental value of $T_2$. A zero-field Rb absorption spectrum at $15^\circ \rm{C}$ is shown in grey. Detuning axis is weighted with respect to the Rb-D2 lines. The table shows the mean parameter values obtained from fits of five spectra. The ENBW, FWHM and FOM values stated also account for the systematic errors involved in linearization. These cascaded-cell filters have both an improved FOM and better spectral profile than single-cell filters.}

\label{fig:wing_filter}
\end{figure*}

\indent Magneto-optical effects can be used to probe all kinds of matter~\cite{das, sargsyan,carr} from non-invasive magnetometry of livestock~\cite{cow} to black hole accretion flows~\cite{blackhole} and vacuum birefringence~\cite{vacuum}. Atomic line filtering is an advantageous magneto-optical bandpass technique owing to its high transmission,  polarization sensitivity and tunability~\cite{higgins}. Applications vary widely including weak signal detection~\cite{pan}, quantum information processing~\cite{sun,eit}, self-stabilizing laser systems ~\cite{LaserStab,luo2021,luo2018}, atmospheric~\cite{LIDAR} and ocean temperature measurements~\cite{ocean2}. Single cell Faraday filters, where a magnetic field is exerted parallel to the k-vector of the light, are discussed widely in the literature~\cite{Faraday1, Faraday3, Faraday4, Faraday5} in particular in rubidium vapor~\cite{Faraday9R, Faraday10R,zielinska}. Spectroscopy in the Voigt geometry, with a magnetic field perpendicular to the k-vector of the light, is less explored~\cite{Voigt4,Voigt5} though several single cell Voigt filters have been built~\cite{Voigt1F,Voigt2F,Voigt3F}. Dependent on the application, a filter can be ‘line center' where filter transmission occurs at the center of the atomic resonance or a ‘wing' type where transmission is detuned from center~\cite{kevfilter}. 
\begin{figure}[!tb]
\centering
{\includegraphics[width=\linewidth]{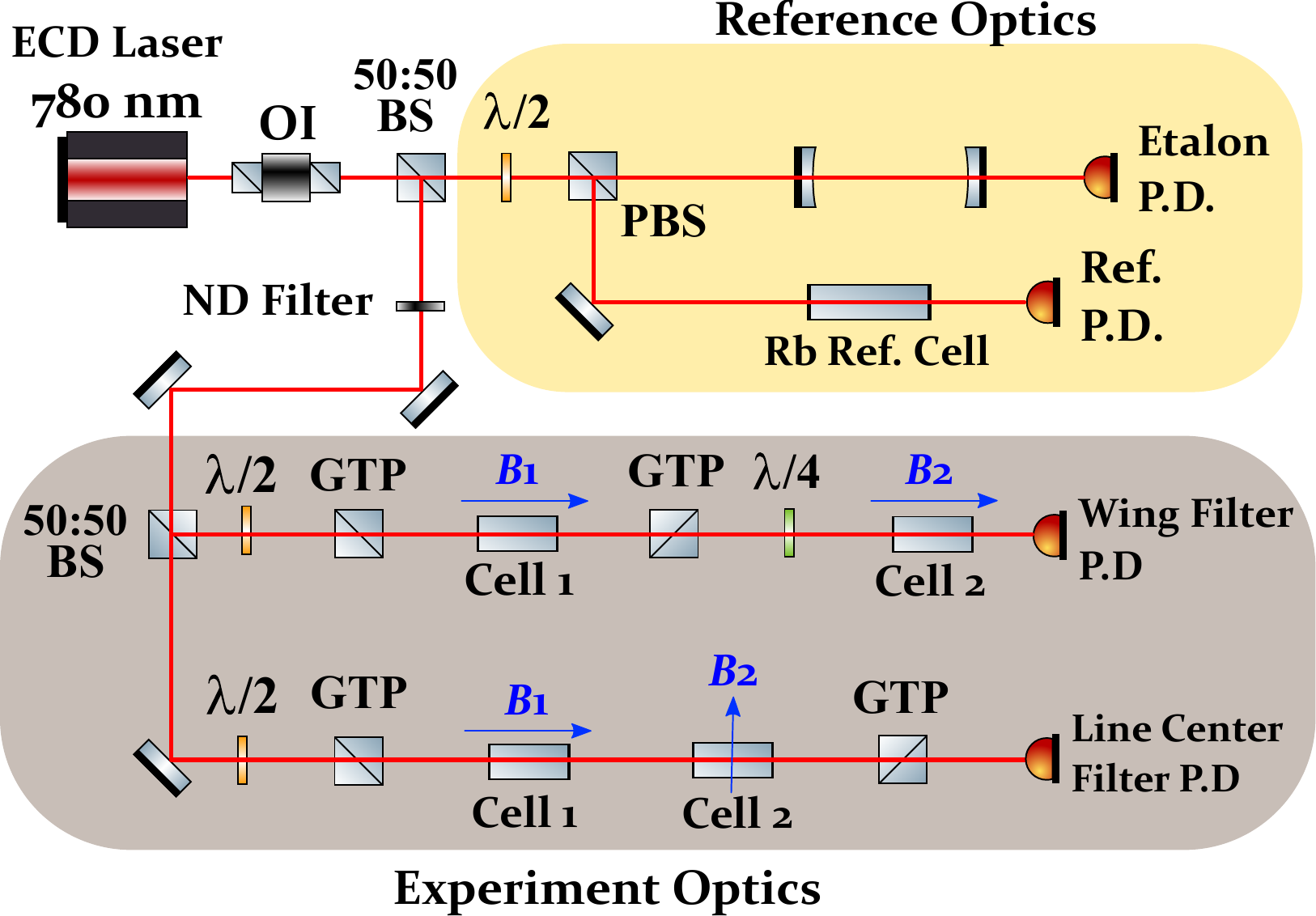}}
\caption{A schematic of the experimental setup.  Light from an external cavity diode (ECD) laser on the Rb-D2 line passes through an optical isolator (OI) and is divided into two paths: reference  and experiment optics. The laser power is attenuated with a neutral density (ND) filter. The first cell in both experiments is placed between crossed Glan-Taylor polarizers (GTP) with an axial magnetic field generated by a solenoid. The solenoid also heats the atoms to reach the required number density. The second cell is placed in either a transverse (line center) or axial (wing) magnetic field. The second cell rests in a separate copper heater. In the wing filter experiment there is a quarter waveplate before the second cell, whereas in the line center experiment, the second cell is placed before the second GTP. We detect output signals with photodetectors (P.D). PBS - Polarizing Beamsplitter, 50:50 BS - 50:50 Beamsplitter}
\label{fig:setup}
\end{figure}
Cascaded wing Faraday-Faraday setups are employed extensively in solar filter setups~\cite{solar1,solar6,solar3} which typically exploit magnetic fields on the order of 1~kG. While magnetic fields homogeneous over the length scale of vapor cells at this magnitude have been realized~\cite{dani,trenec}, high performance Faraday-Faraday filters in low fields have not yet been presented. Voigt-Voigt and Voigt-Faraday filters have been presented in~\cite{solar4} but to our knowledge a Faraday-Voigt configuration has not been discussed in the literature previously.\\
\setlength{\belowcaptionskip}{-6pt}
\begin{figure*}[h]
\centering
{\includegraphics[width=2\columnwidth]{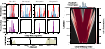}}
\caption{A Rb-D2 Faraday-Faraday wing filter output (purple) for four magnetic fields across the second cell a) Zero field, b) $373$~G, c) $747$~G, d) $2000$~G. Fixed parameters are $T_1~=~86^\circ \rm{C},\hspace{0.05cm} \textit{B}_1=49\hspace{0.05cm}\rm{G},\hspace{0.05cm} \textit{T}_2=110^\circ \rm{C}$ with cell lengths 75~mm and 5~mm. In red, the transmission through the second cell given left hand circular light input. In blue, the filter output if the second cell is removed. In olive, the evolution of the figure of merit (FOM) with second cell magnetic field. The heat map shows the transmission through the second cell given left hand circular light with evolving second cell magnetic field. We experimentally realize the filter with the parameters shown in (c) (see Fig.~\ref{fig:wing_filter}).}
\label{fig:theory_wingfilter}
\end{figure*}
\indent In this Letter, we demonstrate improved wing and line center filter performance on the Rb-D2 line by adding a second cell which absorbs light from the first cell in unwanted transmission regions. We theoretically compute parameters using a modified version of \textit{ElecSus}~\cite{elecsus1,elecsus2} and experimentally realize a Faraday-Faraday wing filter and a Faraday-Voigt line center filter which show excellent agreement with theory. The former is an improvement on previous Rb wing filters and the latter is the largest figure of merit atomic line filter realized to date.
\begin{figure*}[h!]
\centering
{\includegraphics[width=2\columnwidth]{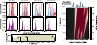}}
\caption{A Rb-D2 Faraday-Voigt line center filter output (purple) for four magnetic fields across the second cell e) Zero field, f) $1254$~G, g) $2528$~G, h) $3000$~G. Fixed parameters are $T_1=100^\circ \rm{C},\hspace{0.05cm} \textit{B}_1 = 162\hspace{0.05cm}\rm{G},\hspace{0.05cm} \textit{T}_2=121^\circ \rm{C}$ with cell lengths 5~mm and 5~mm. In red, the transmission through the second cell given vertical light input. In blue, the filter output if the second cell is removed. In olive, the evolution of FOM with second cell magnetic field. The heat map shows the transmission through the second cell given vertical light input with evolving second cell magnetic field. We experimentally realize the filter with the parameters shown in (g) (see Fig.~\ref{fig:wing_filter}). This cascaded-cell filter  has both an improved FOM and better spectral profile than single-cell filters.}
\label{fig:theory_centreline}
\end{figure*}
\\
\indent We use a figure of merit (FOM) to evaluate filter performance first introduced in~\cite{Ilja}. $\rm{FOM} =$ $\mathcal{T}({\nu}_{\rm{s}})/\rm{ENBW}$ where $\mathcal{T}({\nu}_{\rm{s}})$ is the transmission of the signal frequency, $\nu_{\rm{s}}$. The equivalent noise bandwidth is defined as $\rm{ENBW}=\int$$\mathcal{T}({\nu})$$\rm{d}$${\nu}$/$\mathcal{T}({\nu}_{\rm{s}})$ where ${\nu}$ is the optical frequency. Our figure of merit seeks to maximize the transmission at the signal frequency while minimizing the equivalent noise bandwidth. Optimizations with natural abundance Rb in \textit{ElecSus} show that Faraday-Voigt is the highest figure of merit configuration whereas a Faraday-Faraday configuration yields the best wing lineshapes.
\\
\indent The schematic of our setup is shown in Fig~\ref{fig:setup}. Light scanning over the Rb-D2 line is directed into two experiments in a linear horizontal polarization at a laser power on the order of 100\,nW with a $1/\rm{e}^2$ width of 100\,${\upmu}$m. This ensures we remain in the weak probe regime which our model, \textit{ElecSus}, assumes~\cite{weak}. In both experiments, the first vapor cell is placed after the first Glan-Taylor polarizer with a B-field parallel to the k-vector of the light (Faraday). In the wing filter setup, a second crossed polarizer and a quarter waveplate follows which transforms the linear output light into left hand circular light. This light is input into the second cell, also in the Faraday geometry, before being detected. In the line center experiment, the light output from the first cell is  directed into the second cell before the second polarizer with a magnetic field directed perpendicular to the light's k-vector (Voigt). Part of the light is directed towards a room temperature zero field Rb reference and a Fabry-P\'erot etalon which allows us to calibrate the frequency axis.\\
\indent The first cell's role is to optically rotate the  linearly polarized light while the dominant role of the second cell is to absorb unwanted transmission regions. The angle between the magnetic field and the light k-vector determines the selection rules of the atom-light interaction and the frequencies where light of a particular polarization will be most absorbed~\cite{rotondaro,1970}. In addition, parameters such as temperature and cell length increase the number of atoms the light interacts with thus increasing the strength of transitions induced. In the Faraday geometry, $\sigma^+$/$\sigma^-$ transitions are induced by left/right hand circular light respectively~\cite{adams2018}. By applying larger magnetic fields, the  $\sigma^+$/$\sigma^-$ transition frequencies experience a positive/negative Zeeman shift away from detuning center. At sufficient temperatures, the Doppler widths of the transitions create ‘well-like' lineshapes that absorb over a wider frequency range. In the wing filter horizontal linearly polarized light is input and the vertical component of the rotated light is transmitted by the second polarizer. After traversing a quarter waveplate this light induces $\sigma^+$ transitions in the second vapor cell resulting in significant absorption in the positive detuning region. This allows us to select for the wing in the negative detuning region. Fig.~\ref{fig:theory_wingfilter} and Fig.~\ref{fig:wing_filter} show how the wing filter output varies with magnetic field and temperature across the second cell respectively.
\\
\indent In the Voigt geometry, both $\sigma^+$ and $\sigma^-$ transitions are induced by vertical linearly polarized light. In the line center experiment, the first cell rotates the light from a horizontal to a vertical state. The magnetic field is chosen such that the $\sigma^+$ and $\sigma^-$ absorption wells are shifted leaving a small transmission region around detuning center. This results in high transmission at detuning center and high absorption everywhere else. Fig.~\ref{fig:theory_centreline} and Fig.~\ref{fig:wing_filter} show how the line center filter output varies with magnetic field and  temperature across the second cell respectively.
\\
\indent We use \textit{ElecSus}~\cite{elecsus1,elecsus2} to choose suitable parameters and experimentally verify these predictions for natural abundance rubidium vapor cells. For the wing filter/line center experiment we choose a 75~mm/5~mm first cell placed inside a solenoid. For both experiments the magnetic field across the 5~mm second cell is generated by two NeFeB top hat permanent magnets~\cite{kevfilter} placed in either the Faraday or Voigt geometry. The transverse and axial field over the optical path length is homogeneous to 1\%. We fit the data to our model which show excellent agreement \cite{Hughes2010} with RMS fit errors of 0.6\%/0.09\% for the wing and line center filters respectively. The mean parameters obtained and fits are shown in Fig. \ref{fig:wing_filter}. The wing filter FOM of 0.86~GHz$^{-1}$ is larger than any previously realized Rb wing type filter and the line center filter FOM of 1.63 ~GHz$^{-1}$  is the largest  published to date for an atomic line filter. 
\\
\indent In conclusion, we have shown that dual cell cascaded Rb filters show improvement over the single cell case both in terms of increased FOM and in creating lineshapes that better meet the basic criteria for their applications. We have shown theoretically that this relies on the first and second cells being dominant optical rotators and absorbers respectively. This theory is general and holds for other alkali metals given large enough second cell magnetic fields and temperatures to create the well-like lineshapes. Adding another cell to a setup is an inexpensive and non-intensive step provided the application is not too sensitive to the additional light loss. We plan to give a detailed treatise on the atom-light interactions involved \cite{JB} in a future publication.

\begin{backmatter}
\bmsection{Funding} ESPRC Grant EP/R002061/1

\bmsection{Acknowledgments} We would like to thank Stephen Lishman and the mechanical workshop for fabricating the 5mm heater used. We appreciate the helpful feedback provided by Clare Higgins while drafting this paper and Thomas Cutler and Matthew Jamieson for their advice.

\smallskip

\bmsection{Disclosures} The authors declare no conflicts of interest.

\bmsection{Data availability} Data underlying the results presented in this paper are available in Ref. \cite{data}.

\end{backmatter}

\bibliography{main}


\end{document}